\documentclass[floats]{iucr}              % DO NOT DELETE THIS LINE
\RequirePackage[dvips]{graphicx}
\usepackage{graphicx}% Include figure files
\usepackage{epsfig}% Include figure files
\usepackage{dcolumn}% Align table columns on decimal point
\usepackage{bm}% bold math

\newcommand{\alf}{$\alpha$-AlF$_3$}
\newcommand{\bivo}{$\alpha$-Bi$_4$V$_2$O$_{11}$}
\newcommand{\bivtio}{$\gamma$-Bi$_4$V$_{1.7}$Ti$_{0.3}$O$_{10.85}$}

\newcommand{\RAA}{\AA$^{-1}$}
\newcommand\Qmax{$Q_{max}$}

\newcommand{\etal}{{\it et al.}}

     \paperprodcode{a000000}      % Replace with production code if known
     \paperref{xx9999}            % Replace xx9999 with reference code if known
     \papertype{FA}               % Indicate type of article
     \paperlang{english}          % Can be english, french, german or russian
     \journalcode{J}              % Indicate the journal to which submitted
     \journalyr{xxxx}
     \journaliss{x}
     \journalvol{xx}
     \journalfirstpage{000}
     \journallastpage{000}
     \journalreceived{0 XXXXXXX 0000}
     \journalaccepted{0 XXXXXXX 0000}
     \journalonline{0 XXXXXXX 0000}

\begin{document}                  % DO NOT DELETE THIS LINE

     %-------------------------------------------------------------------------
     % The introductory (header) part of the paper
     %-------------------------------------------------------------------------

     % The title of the paper. Use \shorttitle to indicate an abbreviated title
     % for use in running heads (you will need to uncomment it).

\title{Rapid Acquisition Pair Distribution Function (RA-PDF) Analysis}

\shorttitle{Rapid Acquisition Pair Distribution Function Analysis}

     % Authors' names and addresses. Use \cauthor for the main (contact) author.
     % Use \author for all other authors. Use \aff for authors' affiliations.
     % Use lower-case letters in square brackets to link authors to their
     % affiliations; if there is only one affiliation address, remove the [a].

\author[a]{Peter J.}{Chupas}
\author[b]{Xiangyun }{Qiu}
\author[c]{Jonathan C.}{Hanson}
\author[d]{Peter L.}{Lee}
\author[a]{Clare P.}{Grey}
\author[b]{Simon J. L.}{Billinge}

\aff[a]{Department of Chemistry, State University of New York at Stony Brook,
       \city{Stony Brook}, New York, 11794-3400, \country{USA}}

\aff[b]{Department of Physics and Astronomy, Michigan State University,
        \city{East Lansing}, Michigan, 11793, \country{USA}}

\aff[c]{Department of Chemistry, Brookhaven National Laboratory,
        \city{Upton}, New York, 48824-2230, \country{USA}}

\aff[d]{Advanced Photon Source, Argonne National Laboratory, 
        \city{Argonne}, Illinois, 60439-4803, \country{USA}}

     % Use \shortauthor to indicate an abbreviated author list for use in
     % running heads (you will need to uncomment it).

%\shortauthor{Soape, Author and Doe}

     % Use \vita if required to give biographical details (for authors of
     % invited review papers only). Uncomment it.

%\vita{Author's biography}

     % Keywords (required for Journal of Synchrotron Radiation only)
     % Use the \keyword macro for each word or phrase, e.g. 
     % \keyword{X-ray diffraction}\keyword{muscle}


     % PDB and NDB reference codes for structures referenced in the article and
     % deposited with the Protein Data Bank and Nucleic Acids Database (Acta
     % Crystallographic Section D). Repeat for each separate structure e.g
     % \PDBref[dethiobiotin synthetase]{1byi} \NDBref[d(G$_4$CGC$_4$)]{ad0002}

%\PDBref[optional name]{refcode}
%\NDBref[optional name]{refcode}

\maketitle                        % DO NOT DELETE THIS LINE

\begin{synopsis}

High quality, medium resolution (\Qmax\ $\leq 28.5$~\RAA ) atomic pair distribution functions (PDF) are obtained from crystalline materials using an image plate detector.
Good counting statistics and fast data collection make this technique very promising for studying the structures of nanocrystalline and disordered materials. 
\end{synopsis}

\begin{abstract}

An image plate (IP) detector coupled with high energy synchrotron radiation was used for atomic pair distribution function (PDF) analysis, with high probed momentum transfer \Qmax\ $\leq 28.5$~\RAA\ from crystalline materials. 
Materials with different structural complexities were measured to test the validity of the quantitative data analysis.
Experimental results are presented here for crystalline Ni, crystalline \alf , and the layered Aurivillius type oxides \bivo\ and \bivtio .
Overall, the diffraction patterns show good counting statistics with measuring time from one to tens of seconds.  The PDFs obtained are of high quality.
Structures may be refined from these PDFs, and the structural models are consistent with the published literature. Data sets from similar samples are highly reproducible. 

\end{abstract}

     %-------------------------------------------------------------------------
     % The main body of the paper
     %-------------------------------------------------------------------------
     % Now enter the text of the document in multiple \section's, \subsection's
     % and \subsubsection's as required.

\section{Introduction}

Technologically important materials are becoming structurally more complex and disordered, and encompass a mixed range of amorphous, nano-crystalline, and crystalline materials.  Recently there has been a push to characterize the structure of these materials on atomic length scales by using the atomic pair distribution function (PDF) method \cite{egami;b;utbp03}.  
This method makes use of both the diffuse and Bragg scattering components, and proves an attractive method when powder diffraction patterns prove unsatisfactory for the more traditional Rietveld analysis \cite{petko;jacs00,petko;prb02}.
PDF analysis has been a method of choice for amorphous and liquid samples for many years \cite{warre;bk90}. With the advent of advanced synchrotron based X-ray, pulsed neutron sources, and fast computing it has more recently been successfully applied to the study of crystalline materials \cite{egami;mt90,toby;aca92,egami;b;utbp03}.  

The study of nano-crystalline and crystalline materials benefits from higher real-space resolution measurements than are typically used for amorphous or liquid samples \cite{petko;prl00}.  
An important experimental requirement to obtain high real-space resolution in the PDF, is to measure the structure function $S(Q)$ to a high value of scattering vector {\it Q}.  Conventional high real-space resolution  measurements typically make
use of energy resolving point detectors, such as high-purity germanium, that are scanned over
wide angular ranges.  These measurements are very slow and generally take more than eight hours, even at a synchrotron.  
This, coupled with the novelty of the approach and the somewhat intensive
data analysis requirements, has prevented 
widespread application of the technique in areas such  as nano-materials. In the present paper
we show how high-quality medium-high resolution PDFs (\Qmax\ $\leq 28.5$~\RAA ) can be
obtained in a few seconds of data collection time using a two-dimensional (2D) image plate (IP) detector.  The PDFs
were successfully obtained from the data using a widely available data preprocessing computer packages coupled with a home-written data analysis program.  These results open the way for more
widespread application of the PDF technique to study the structure of nanocrystalline materials.  
They also open up the possibility for qualitatively new experiments to be carried out such as time-resolved studies of local structure.

Recent developments have shown the utility of 2D detector technology scattering studies of liquids.  A recent report by Crichton \etal\ (2001), has made use of integrated two-dimensional IP data for {\it in-situ} studies of scattering from liquid GeSe$_2$.  This study, and others more recently, demonstrate the feasibility of using IPs for diffuse scattering measurements though the measurements have been limited in real-space resolution, with \Qmax\ $\leq 13.0$~\RAA~\cite{mezou;rsi02,krame;jncs03}, making them less suitable for the study of crystalline and nanocrystalline materials.
Image plates have also been successfully used to study diffuse scattering from single crystals~\cite{ester;pt98}. 
A Debye-Scherrer camera utilizing IPs has also been tested in one of our groups and shows promise for lower energy x-ray sources such as laboratory and second generation synchrotron sources~\cite{stach;rsi00}. 

Successful application of IP technology to the measurement of quantitatively
reliable high real-space resolution PDFs requires that a number of issues are resolved.  For example, it is necessary to correct for
contamination of the signal from Compton and fluorescence intensities and for angle and energy
dependencies of the IP detection efficiency \cite{zales;jac98,ito;nima91}.  Here we show that high quality medium-high real-space resolution PDFs can
be obtained by applying relatively straightforward  corrections. 
As expected, the quality of the PDFs is lower in samples comprising  predominantly
low atomic-number elements.  Nonetheless, even the PDFs of these samples prove adequate.  Studies are underway to optimize the  experimental setup and our studies to date suggest that it will be possible to study  low-Z materials by  using this approach.

%Since this study represents, to our knowledge, the first use of high energy synchrotron radiation coupled with image plate detection for atomic pair distribution function analysis. 
A series of crystalline materials with differing structural complexities and with a wide range of scattering powers facilitates our investigation of various aspects of this technique.  Here we present results from crystalline Ni, \alf , and the layered Aurivillius type oxides \bivo\ and \bivtio .

\section{Method}

All diffraction experiments were performed at the 1-ID beam line at the Advanced Photon Source (APS) located at Argonne National Laboratory, Argonne, IL (USA).
High energy X-rays were delivered to the experimental hutch using a double bent Laue monochromator capable of providing a flux of 10\(^{12}\) photons/second and operating with X-rays in the energy range of 80-100~keV \cite{shast;jsr02}.
Two energies of X-rays were used for the experiments, 80.725~keV (0.15359~\AA ) and 97.572~keV (0.12707~\AA ), with the 80.725 keV experiments being performed first.  
In the optics hutch, a gold foil was installed after the monochromator between two ion chambers, attenuating the flux of the beam by approximately 30.0 percent.  Calibration of energy at 80.725~keV was achieved using the gold absorption edge as the reference.

A Mar345 image plate camera, with a usable diameter of 345 mm, was mounted orthogonal to the beam path, with the beam centered on the IP.  When operating with 80.725~keV X-rays, a LaB$_6$ standard was used to calibrate the sample to detector distance and the tilt of the IP relative to the beam path, using the software Fit2D \cite{hamme;hpr96,hamme;esrf98}.  
For calibration of the IP, the wavelength was fixed to represent the gold absorption edge, 0.15359~\AA , at 80.725~keV.  When the X-ray energy was increased from 80.725 to 97.572~keV, the sample to detector distance was first fixed at the value determined at the Au edge and the wavelength was calibrated using the standard LaB$_6$. The IP camera was then moved closer to the sample and the new sample to detector distance was obtained from refinement by fixing the wavelength at 0.12707~\AA.
Sample to detector distances of 317.28 and 242.12~mm were used for collection of data with measured \Qmax\ of 21.0~\RAA\ and 30.0~\RAA , respectively. 
In practice, \Qmax\ of 18.5~\RAA\ and 28.5~\RAA\ were used due to the corrupted data near the IP edges.

\begin{figure}\label{fig;niIP2DIq}
\caption{(color) Two dimensional contour plot from the Mar345 Image Plate Detector.  The data are from nickel powder measured at room temperature with 97.572~keV incident X-rays. The concentric circles are where Debye-Scherrer cones intersect the area detector.}
\includegraphics[height=3.4in]{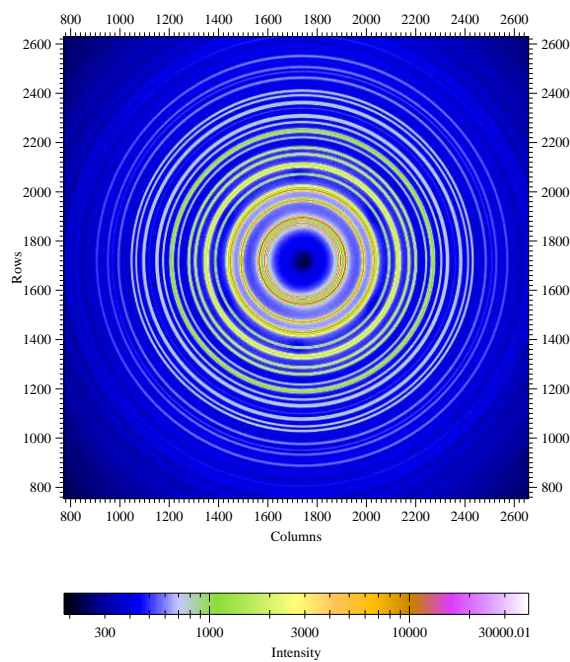}
\end{figure}

The methods used to synthesize \alf , \bivo\ and \bivtio\  have been reported elsewhere~\cite{chupa;jacs01,kim;s02}. Ni was purchased from Alfa Aesar (99.9\%, 300 mesh) and was used as received. 
Fine powders of all the samples were measured in flat plate transmission geometry, with thickness of  1.3~mm packed between kapton foils. 
The beam size on the sample as defined by the final slits before the goniometer was 0.4~mm $\times$ 0.4~mm.  Lead shielding before the goniometer, with a small opening for the incident beam, was used to reduce background.  
All raw data were integrated using the software Fit2D and converted to intensity versus $2\theta$ (the angle between incident and scattered X-rays).  An example of the data from nickel measured at 97.572~keV is shown in Fig.~\ref{fig;niIP2DIq}. 
The integrated data were then transferred to a home-written program, RAPDFgetX (unpublished), to obtain the PDF.  

\begin{figure}\label{fig;nifqgr}
\caption{
%(a)Experimentally obtained structure function $S(Q)$,
%         (b)Pair distribution function $G(r)$,
         (a) the experimental reduced structure function $F(Q)=Q*(S(Q)-1)$ of Ni powder. The inset is a zoom in of high {\it Q} region showing the excellent signal to noise.
         (b) The experimental $G(r)$ (solid dots) and the calculated PDF from refined structural model (solid line). The difference curve is shown offset below.}
\includegraphics[width=2.6in]{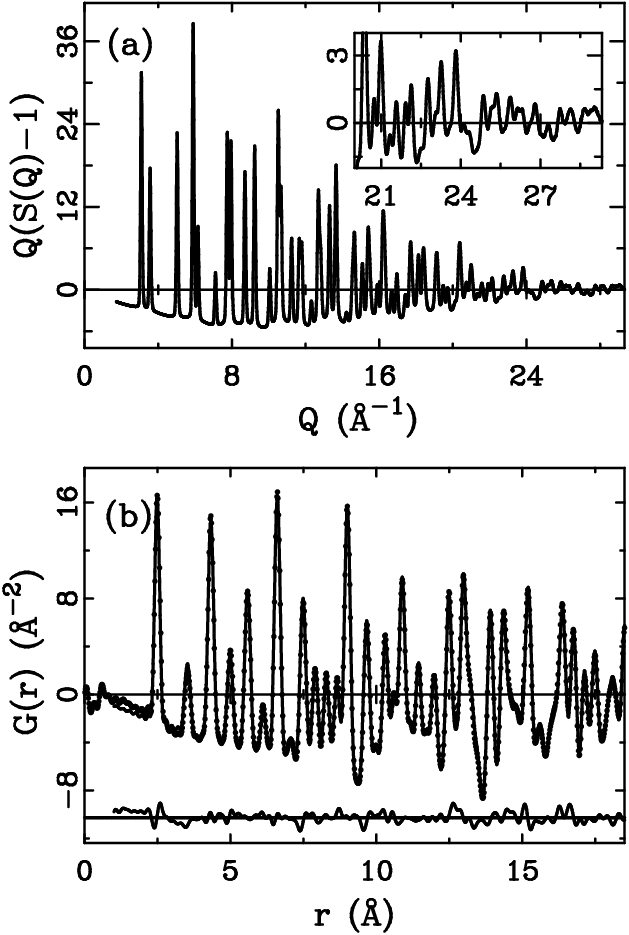}
\end{figure}

Distortions in intensities when using flat IPs have been addressed in single crystal crystallography, and arise from the fact that IPs are of a finite thickness, often between 100-200 microns \cite{zales;jac98,wu;jac02}.
%(Gruner, 1993; Zalenski et al., 1998; Wu et al., 2002).
At the high energies ($\geq$ 60 keV) needed to measure $S(Q)$ to a high value of {\it Q} with commercial IP cameras, absorption of the X-ray photons by the phosphor is very small 
and most of the X-rays travel straight through the phosphor (the thin phosphor regime).  The absorption, and thus the measured intensity, is then dependent on the path length of the beam through the IP phosphor at all incident angles and can easily be corrected. \cite{zales;jac98}.  This oblique incidence correction assumes that the correction is equal in all directions, thus, care was taken to ensure the IP was mounted orthogonal to the beam.  Standard corrections for multiple scattering, polarization, absorption, Compton scattering, and Laue diffuse scattering were also applied to the integrated data to obtain the reduced structure function $F(Q)$ (Fig.~\ref{fig;nifqgr} (a)).
Direct Fourier transformation gives the pair distribution function $G(r)$ (Fig.~\ref{fig;nifqgr}(b)).
The average structure models were refined using the profile fitting least-squares regression program, PDFFIT \cite{proff;jac99}.  Rietveld refinements of the data were performed with GSAS~\cite{larso;unpub87}.

\bigskip

%
%%%%%%%%%%%%%%%%%%%%%%%%%%%%%%%%%%%%%%%%%%%%%%%%%%%%%%%%%%%%%%%%%%%
%  Float starts here
%%%%%%%%%%%%%%%%%%%%%%%%%%%%%%%%%%%%%%%%%%%%%%%%%%%%%%%%%%%%%%%%%%%
%%%%%%%%%%%%%%%%%%%%%%%%%%%%%%%%%%%%%%%%%%%%%%%%%%%%%%%%%%%%%%%%%%%
%  Float stops here
%%%%%%%%%%%%%%%%%%%%%%%%%%%%%%%%%%%%%%%%%%%%%%%%%%%%%%%%%%%%%%%%%%%%
%
%%%%%%%%%%%%%%%%%%%%%%%%%%%%%%%%%%%%%%%%%%%%%%%%%%%%%%%%%%%%%%%%%%%
%  Float starts here
%%%%%%%%%%%%%%%%%%%%%%%%%%%%%%%%%%%%%%%%%%%%%%%%%%%%%%%%%%%%%%%%%%%
%%%%%%%%%%%%%%%%%%%%%%%%%%%%%%%%%%%%%%%%%%%%%%%%%%%%%%%%%%%%%%%%%%%
%  Float stops here
%%%%%%%%%%%%%%%%%%%%%%%%%%%%%%%%%%%%%%%%%%%%%%%%%%%%%%%%%%%%%%%%%%%%

\section{Results} 

All the experimental data collected from samples of Ni, \alf , \bivo\ and \bivtio\ show excellent counting statistics over the entire {\it Q} range, reflecting the significant advantage obtained by extracting 1D data sets by integration of a 2D area detector.
Image plate exposure time, ranging from one second (Ni) to 10 seconds (\alf ), was chosen carefully for each sample to maximize the dynamic range of the IP without saturating the phosphor on the Bragg peaks.  All the PDFs, $G(r)$s, show either superior or acceptable qualities. 
% based on the quality criteria suggested by Peterson \etal\ (2003).
%

The first example, shown in Fig.~\ref{fig;nifqgr}, is from standard Ni powder with \Qmax\ of 28.5~\RAA .  
The Ni PDF, $G(r)$, in Fig.~\ref{fig;nifqgr} (b) appears to have minimal systematic errors which appear as the small ripples before the first PDF peak at $r=2.4$~\AA . These result from imperfect data corrections and their small amplitude is a good indication of the high quality of the data.
A structural model (space group $Fm\overline{3}m$) was readily refined, and gave excellent agreement with the data
as shown in Fig.~\ref{fig;nifqgr}(b). 
The lattice parameter (3.5346(2)~\AA ) and an isotropic thermal displacement parameter ($U = 0.005184(6)$~\AA$^2$) were refined.  The lattice parameters reproduce the expected values given in previously published data~\cite{billi;b;lsfd98}. In spite of the simplicity of the Ni crystal structure, the exceptional quality of both the experimental Ni PDF and refinement indicates that the necessary data corrections of image plate data with high {\it Q} range (28.5~\RAA\ in this case) can be carried out properly and with an acceptable level of accuracy.

\begin{figure}
\label{fig;alf3fqgr}
\caption{(a) Experimental reduced structure function $F(Q)=Q*(S(Q)-1)$ of \alf ,
         (b) $G(r)$ and modeled PDF of \alf, notations as in Fig. ~\ref{fig;nifqgr}.}
\includegraphics[width=2.6in]{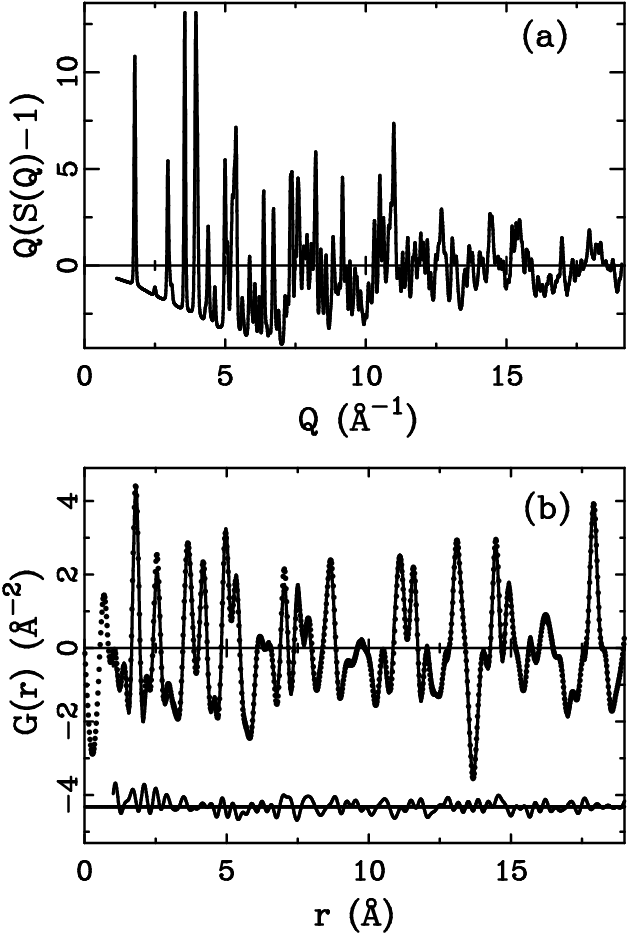}
\end{figure}

The weakly scattering \alf\ compound, in addition to a slightly more complex structure than Ni, presents a greater challenge to proper data correction due to the majority contribution of Compton scattering in the  high {\it Q} region.
The reduced structure function and resulting PDF are shown in Fig.~\ref{fig;alf3fqgr}.
The data were successfully refined using a previously reported model from the literature (space group $R\overline{3}c$) \cite{danie;jpcm90}.
The fit is shown in Fig.~\ref{fig;alf3fqgr}(b).  The overall quality of the fit is good with a weighted profile agreement factor~\cite{proff;jac99} of 3 percent. 
The refined lattice parameters are $a=b=4.9420(3)$~\AA , $c=12.4365(2)$~\AA .  The fractional coordinate of the F atom is $x=0.4267(4)$.  These numbers are consistent with the Rietveld refinement of the same data, which produced refined lattice parameters of $a=b=4.9383(4)$~\AA\ and $c=12.4271(1)$~\AA , and a fractional coordinate, $x=0.4287(4)$, for the F atom.
Both refinements are consistent with the published data of 4.9381(5)~\AA , 12.4240(3)~\AA , and 0.4309(4), respectively \cite{chupa;jacs01}.
The successful application of PDF analysis of \alf\ data indicates this technique is capable of handling weakly scattering materials and still gives reliable information.
%%%%%%%%%%%%%%%%%%%%%%%%%%%%%%%%%%%%%%%%%%%%%%%%%%%%%%%%%%%%%%%%%%%
%  Float starts her
%%%%%%%%%%%%%%%%%%%%%%%%%%%%%%%%%%%%%%%%%%%%%%%%%%%%%%%%%%%%%%%%%%%
%%%%%%%%%%%%%%%%%%%%%%%%%%%%%%%%%%%%%%%%%%%%%%%%%%%%%%%%%%%%%%%%%%%
%  Float stops here
%%%%%%%%%%%%%%%%%%%%%%%%%%%%%%%%%%%%%%%%%%%%%%%%%%%%%%%%%%%%%%%%%%%%

\begin{figure}\label{fig;bivo_bivtio}
\caption{ (a)  Reduced structure function F({\it Q}) of \bivo ,
          (b) Experimentally obtained $G(r)$'s of \bivo\ (solid dots) and \bivtio\ (solid line). The differences between them are plotted below with an offset, noting the high reproducibility at low {\it r} region (also shown in the inset). }
\includegraphics[width=2.8in]{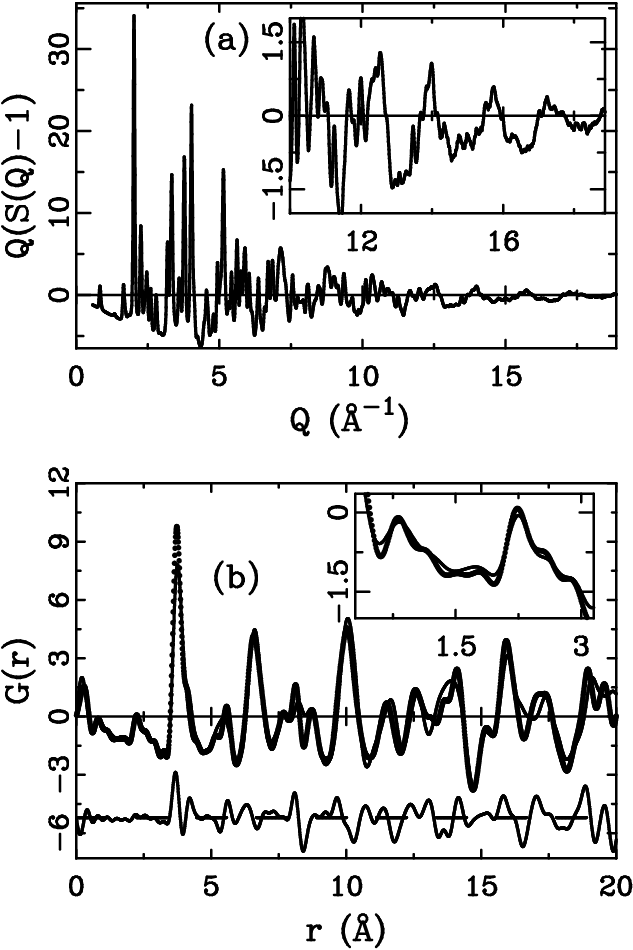}
\end{figure}

In addition to the highly crystalline and rather simple nickel and \alf\ structures, the more complex and disordered layered Aurivillius type oxide anion conductors \bivo\ and \bivtio\ were examined.  
The structure is derived from the ordered Bi$_4$Mo$_2$O$_{12}$\ structure, which contains alternating Bi$_2$O$_2^{2+}$\ layers spaced by corner sharing perovskite MoO$_6$\ layers~\cite{kim;s02,yan;ssi95}.  
The vanadium analogues contain significant disorder in the vanadium layers due to oxygen vacancies.  
Vanadium can exist in 4, 5, and 6 coordinate environments, inducing significant static disorder in these layers, which is further complicated by dynamic disorder resulting from the high anionic conductivity of these materials.  
Titanium substitution on the vanadium sites leads to the stabilization of the high temperature $\gamma$-phase. The experimentally obtained reduced structure function, F({\it Q}), of \bivo\ is shown in Fig.~\ref{fig;bivo_bivtio}(a). A combined PDF and Rietveld analysis was carried out on the same data set and detailed results will be reported elsewhere. 
To examine the reproducibility of the systematic errors in the IP data analysis, the measured data for the compounds \bivo\ and \bivtio\ were compared (Fig.~\ref{fig;bivo_bivtio}(b)).
% compares the experimental $G(r)$'s between \bivo\ and \bivtio , with the difference plotted below. 
%Systematic errors arising from various imperfect corrections are almost inevitable, and they are mostly slowly varying functions which give ripples in the unphysical low-$r$ region.
Clearly, the ripples in the low $r$ regions are rather small and highly reproducible, implying the high reliability of the PDF data especially for comparative studies.
For example, very small differences exist around the first V-O bond length of 1.80~\AA , as evident in the inset of Fig.~\ref{fig;bivo_bivtio}(b). 
The significant differences beyond the first peak reveal the somewhat drastic structural changes on longer length scales. This also shows that RA-PDF is capable of capturing structural changes where they exist, reproducing local structures if unchanged. 
% 
%%%%%%%%%%%%%%%%%%%%%%%%%%%%%%%%%%%%%%%%%%%%%%%%%%%%%%%%%%%%%%%%%%%
%  Float starts here
%%%%%%%%%%%%%%%%%%%%%%%%%%%%%%%%%%%%%%%%%%%%%%%%%%%%%%%%%%%%%%%%%%%
%%%%%%%%%%%%%%%%%%%%%%%%%%%%%%%%%%%%%%%%%%%%%%%%%%%%%%%%%%%%%%%%%%%
%  Float stops here
%%%%%%%%%%%%%%%%%%%%%%%%%%%%%%%%%%%%%%%%%%%%%%%%%%%%%%%%%%%%%%%%%%%% 
%
%%%%%%%%%%%%%%%%%%%%%%%%%%%%%%%%%%%%%%%%%%%%%%%%%%%%%%%%%%%%%%%%%%%
%  Float starts here
%%%%%%%%%%%%%%%%%%%%%%%%%%%%%%%%%%%%%%%%%%%%%%%%%%%%%%%%%%%%%%%%%%%
%\begin{figure}\label{fig;bisq}
%\caption{(a) Reduced structure function F({\it Q}) of \bivo\ , 
%         (b) Reduced structure function F({\it Q}) of \bivtio\
%}
%\includegraphics[width=2.6in]{bivo315fqgrfit.ps}
%\end{figure}
%%%%%%%%%%%%%%%%%%%%%%%%%%%%%%%%%%%%%%%%%%%%%%%%%%%%%%%%%%%%%%%%%%%
%  Float stops here
%%%%%%%%%%%%%%%%%%%%%%%%%%%%%%%%%%%%%%%%%%%%%%%%%%%%%%%%%%%%%%%%%%%%  
%

\section{Discussion}

Previous image plate measurements of diffuse scattering have been limited to a \Qmax\ of 13~\RAA .  In the case of the RA-PDF experiment a much higher {\it Q} range (\Qmax\ of 30~\RAA ) is measured which is necessary for PDF analysis of crystalline and nanocrystalline compounds.  Here we discuss some of the problems inherent with the use of 2D detectors for PDF data collection.  The good quality of the PDFs presented in this paper show that these problems are relatively minor, though they will be explicitly addressed in future experiments.  The data from low-Z materials are more problematic to analyze and we have not presented any PDFs from a low-Z material measured above $Q_{max}=18.0$~\RAA .  We expect this can be resolved by addressing the problems described here.

The application of an IP camera (where readout and bleaching procedures are integrated) in these studies was chosen for two reasons.  First, utilization of an IP in a flat geometry was chosen to simplify the data correction procedure.  Second, the ability to read out images in rapid succession facilitates the future implementation of time resolved experiments.  Thus, it was necessary to condense the diffraction pattern into the usable IP area by moving to high incident X-ray energies and minimizing sample to IP distances. 
Our IP diffraction data have only moderate {\it Q} space resolution (0.05~\RAA ).  This results in lost structural information in the high-$r$ region, for example beyond $r=40.0$~\AA\ in the Ni PDF.
The majority of the loss in {\it Q}-resolution comes from the finite sample size effects and could be improved by reducing the gauge volume by reducing beam size and sample thickness.
Even though it may not be feasible to achieve the {\it Q}-resolution of conventional PDF measurements ($\ge\ 0.02$~\RAA ), the current medium {\it Q}-resolution does not present a serious problem in applications involving a wide range of complex structures of current interest. As we discussed, the primary effect of the moderate resolution is the loss of high-$r$ information in the PDF. This high-$r$ information is often neglected in analysis of PDF data.    

Pushing the accessible {\it Q} range of the image plate also introduces other problematic side effects.
The lack of energy resolution of the phosphor in image plates causes the inelastic component to be measured over the entire measured {\it Q} range.
In conventional measurements using energy resolving detectors the Compton and fluorescence scattering can be resolved from the elastic component.
The coherent elastic scattering contains the structure information of interest but its intensity drops quickly with increasing {\it Q}. 
The Compton intensity can be corrected using tabulated parameters. 
However, the incoherent intensity increases with {\it Q}, and can be as much as five times as strong as the coherent intensity,  even around {\it Q}$=25.0$~\RAA\ \cite{petko;prl00}.
This results in signal-to-noise problems, and also a small aberration to this correction can result in a large distortion to the extracted elastic intensity.
In the current correction scheme the change in energy of the inelastically
scattered Compton intensity is not explicitly included.  This affects the measured intensity due to the energy dependence of absorption and detector efficiency corrections.  
These are expected to be fairly small since $\Delta E/E \approx 2.1\%$ at $2\theta = 30.0$ with an X-ray energy of 80~keV. 
Imperfect Compton corrections will have a more significant effect in low-Z materials where the Compton cross-section is dominant.  However, the current results indicate that even with the low-Z \alf\ material the Compton scattering can be adequately corrected, at least up to 18~\RAA .

Data collection with image plates can be subject to quite high backgrounds, especially in experiments utilizing high energy X-rays.  For example, the detection area of an IP records all air scattering from the direct beam, between the final scattering slits to the beam stop.  In conventional angle dispersive measurements collimation after the sample position confines the visible air scattering to be the small area around the sample. 
It is extremely important to reduce the background scattering for quantitative PDF studies and implementations such as extensive lead shielding around the beam path, to minimize parasitic scattering, and collimation, to decrease air scatter, should be used.
Despite the non-optimized setup, the preliminary results shown here indicate quantitatively that high quality PDFs from crystalline materials are possible using this approach, opening the way to novel experiments such as time-resolved PDF measurements.
The problems addressed above will be further investigated in future measurements. 
%Regardless, without the implementation of fully characterized corrections, the systematic errors evident in the PDF at low-$r$ are at an acceptable level and the measurements demonstrate a high level of reproducibility.

%
%
%
%

\section{Conclusions}

Medium-high real-space resolution PDF data analysis from crystalline materials was performed by using image plate data and shows promising results. 
Comparable or even better statistics than from conventional X-ray measurements can be achieved with significantly shorter counting times. 
The new combination of a real space probe and fast counting time opens up a broad field for future applications to wide variety of materials of both scientific and technological interest. For example, PDF methods could be used to study of structural changes under in situ conditions~\cite{norby;ct98} and the time development of chemical reactions and biological systems over short time scales of seconds may be studied. 
%Further investigations of various aspects of this method are under way.   
%\subsection{Title}%Text text text text text text text text text text text text text text%\subsubsection{Title}

\ack{Acknowledgments}

This work is supported by National Science Foundation through grants DMR-0075149, DMR-0211353 and CHE-0211029, also by U.S. Department of Energy through grant DE-FG02-97ER45651, DE-FG02-96ER14681 and DE-AC02-98CH10086.  John Palumbo and Namjun Kim are thanked for their help with data collection and for sample synthesis, respectively. The Advanced Photon Source is supported by the U.S. Department of Energy through grant W-31-109-ENG-38.

     % References are at the end of the document, between \begin{references}
     % and \end{references} tags. Each reference is in a \reference entry.

\end{document}